\documentclass{PoS}

\usepackage{cancel}

\newcommand{\dd}{\mathrm{d}}

\newcommand{\bbm}{\left(\begin{matrix}}
\newcommand{\ebm}{\end{matrix}\right)}
\newcommand{\beq}{\begin{eqnarray}}
\newcommand{\eeq}{\end{eqnarray}}

\newcommand{\cbral}{[\![}
\newcommand{\cbrar}{]\!]}

\newcommand{\sfrac}[2]{{\textstyle\frac{#1}{#2}}}

\newcommand{\be}{\begin{equation}}
\newcommand{\ee}{\end{equation}}

\newcommand{\beqa}{\begin{eqnarray}}
\newcommand{\eeqa}{\end{eqnarray}} 
\def\nn{\nonumber} \def \bea{\begin{eqnarray}} \def\eea{\end{eqnarray}}

\newcommand{\barr}{\begin{array}}
\newcommand{\earr}{\end{array}}
  \def\G{\Gamma}
 \def\d{\delta} 

 \def\L{\Lambda}

   \def\X{\mathbb X} \def \A{\mathbb A}
 \def \B{\mathbb B}
\def \L{\mathbb L}
 \def\one{\mbox{1 \kern-.59em {\rm l}}}

\def\sfp{{\mathsf p}}
\def\sfL{{\mathsf L}}

\def\bit{\begin{itemize}} \def\eit{\end{itemize}}

\def\({\left(} \def\){\right)}

\title{The Algebroid Structure of Double Field Theory}

\ShortTitle{The Algebroid Structure of DFT}

\author{Athanasios Chatzistavrakidis, Larisa Jonke, \speaker{Fech Scen Khoo}
\\
        Rudjer Bo\v skovi\' c Institute, Zagreb\\
        E-mail: \email{athanasios.chatzistavrakidis@irb.hr}, \email{larisa@irb.hr},  \email{Fech.Scen.Khoo@irb.hr}}

\author{Richard J. Szabo\\
        Heriot-Watt University, Maxwell Institute for Mathematical Sciences, The Higgs Centre for Theoretical Physics, Edinburgh \\
        E-mail: \email{R.J.Szabo@hw.ac.uk}}

\abstract{
By doubling the target space of a canonical Courant algebroid 
and subsequently projecting down to a specific subbundle, 
we identify the data of double field theory (DFT) and hence 
define its algebroid structure.
We specify the properties of the DFT algebroid.
We show that one of the Courant algebroid properties 
plays the role of the strong constraint in the context of DFT.
The DFT algebroid is a special example when 
properties of a Courant algebroid are relaxed
in a specific and dependent manner.
When otherwise, we uncover additional structures.}

\FullConference{Corfu Summer Institute 2018 "School and Workshops on Elementary Particle Physics and Gravity"\\
		(CORFU2018)\\
		31 August - 28 September, 2018\\
		Corfu, Greece}

\begin{document}

\section{Motivation}

The notion of a Courant algebroid as a central structure in generalized geometry
is defined through a set of axioms \cite{C90, LWX, Roy99, Sev, Hitchin, Gualtieri}. 
Its original characterization in Ref. \cite{LWX} is in terms of five axioms, 
however only three out of them form a minimal set that sufficiently quantifies a Courant algebroid \cite{Uchi}.
These properties are basically those of the objects comprising the algebroid structure, namely its bracket, its bilinear form and the anchor map to the tangent bundle of the underlying smooth manifold.
Twistings of the properties of the Courant algebroid have been studied and they were found to lead to relaxed structures, such as pre-Courant algebroids \cite{Vais04,HS09}.

On the other hand, double field theory (DFT) is a T-duality covariant reformulation of supergravity.
As its name suggests, the theory is constructed on a doubled geometry \cite{Siegel_1, Siegel_2, HZ09_1, HZ09_2, HHZ1003, HHZ1006}. 
Therefore, a strong constraint or at times known as a section condition becomes an intrinsic feature of the theory.
Only upon the imposition of the strong constraint, which effectively amounts to eliminating half of the dependence of the fields in the theory on the doubled spacetime,
the theory reduces to supergravity on the physical space.
From the perspective of a mathematical structure, the gauge structure of DFT results in a Courant algebroid after imposing the strong constraint \cite{HZ09_2}. 

Our goal in this contribution is twofold. First, to determine what is the underlying mathematical structure of the doubled geometry before the strong constraint is solved, and how  it is related to geometric structures obtained upon relaxing some of the Courant algebroid axioms. Second, to examine whether such a structure not only gives rise to a Courant algebroid upon solving the strong constraint, but also  originates in some precise sense from a large Courant algebroid on a doubled space.

There exist several proposals in this regard \cite{DSt, DS16, HIW16, FRS17, S18, MS18}.
For instance, DFT formulated in the supermanifold language in the works of \cite{DS16, HIW16}, utilizes the
correspondence between Courant algebroids and QP-manifolds \cite{Roy02}.
The present contribution is based on \cite{CJKS}, where
we found that the standard Courant algebroid language of formulation accomodates a larger class of structures en route to DFT.

\section{Courant algebroid}

In this section we outline the definition of a Courant algebroid 
and its properties.
Let $E \rightarrow M$ be a vector bundle.
For our purposes it is enough to make the simplifying assumption that 
 $E=TM \oplus T^*M$, where $M$ is a $d$-dimensional manifold, which is locally true always.
Sections of the generalized bundle $E$ are generalized vectors
$A=A_V+A_F\in \G(E)$, where $A_V\in\G(TM)$ is a vector field and $A_F\in\G(T^{\ast}M)$ is a one-form.

An anti-symmetric bracket can be defined for the sections of the bundle, $[\cdot,\cdot]_E: \Gamma(E) \otimes \Gamma(E) \rightarrow \Gamma(E)$.
In addition,  there is a non-degenerate symmetric bilinear form, $\langle \cdot , \cdot \rangle_E: \Gamma(E) \otimes \Gamma(E) \rightarrow C^{\infty}(M)$
which takes as input two sections of the bundle and gives a scalar output.
Lastly, there is an anchor map, $\rho: E \rightarrow TM$. 
This quadruple $(E, [\cdot,\cdot]_E, \langle \cdot , \cdot \rangle_E, \rho)$ defines a Courant algebroid \cite{LWX}.

The notion of a Courant algebroid allows for two types of binary operations, the anti-symmetric Courant bracket or the non-anti-symmetric Dorfman bracket \cite{Roy99}.
Anti-symmetrizing the Dorfman bracket yields the Courant bracket, and the corresponding definitions of a Courant algebroid are equivalent.
We will be working with the standard Courant bracket,
which is given by
\bea 
[A,B]_{{\rm s}E}=[A_V,B_V]+{\cal L}_{A_V}B_F-{\cal L}_{B_V}A_F-\sfrac 12\, \dd (\iota_{A_V}B_F-\iota_{B_V}A_F)~.
\eea  
The Courant bracket extends from the first term, which is the usual Lie bracket of vector fields, to include one-form contributions.
It can be further twisted by an additional one-form $H(A_V, B_V)$, where $H$ is a closed three-form.
On the other hand, the Dorfman bracket or Dorfman derivative is
\be 
 A \circ B = L_A B = [A_V,B_V]+{\cal L}_{A_V}B_F - \iota_{B_V} \dd A_F~,
\ee
which is clearly a generalization of the Lie derivative.
For the symmetric pairing
\be 
\langle A,B\rangle_{sE}=  \sfrac 12 (\iota_{A_V} B_F + \iota_{B_V} A_F)~,
\ee
the $O(d,d)$ symmetry is inherent in the structure as
\be 
\langle A,B\rangle_{sE} = \sfrac 12
\left(\begin{array}{cc}
A_V & A_F \\
\end{array}
\right)
\left(\begin{array}{cc}
0 & 1^d  \\
1^d & 0\\
\end{array}
\right)
\left(\begin{array}{cc}
B_V  \\
B_F
\end{array}
\right)
~,
\ee
with {\small 
$\left(\begin{array}{cc}
0 & 1^d  \\
1^d & 0\\
\end{array}
\right)$
}
being the $O(d,d)$ metric which we denote by $\eta_{IJ}$, $I,J$ being $2d$-dimensional indices. Therefore, the pairing is $O(d,d)$ invariant.

There are five main properties that the above structures satisfy.	
First is the (modified) Jacobi identity for the Courant bracket\footnote{From here onwards we remove the subscript $sE$ as we will only discuss the standard Courant algebroid case.},
	\begin{equation}
	\label{mod_Jac}
	[[A,B],C] + cyclic = \sfrac 13\,  {\cal D} \langle [A,B],C\rangle +{cyclic}
	~,
	\end{equation}
		where 
	the differential operator ${\cal D}: C^{\infty}(M)\to \G(E)$ is defined by
	\begin{equation} 
	\langle {\cal D}f, A\rangle=\sfrac 12\,\rho(A)f~,
	\end{equation}
for any 
	$A,B,C\in\G(E)$ and $f\in C^{\infty}(M)$. Note that for the Dorfman bracket the exact term on the right-hand side of (\ref{mod_Jac}) is absent; the corresponding Jacobi identity is in Leibniz-Loday form however, since the operation is not anti-symmetric.
Next property is the (modified) Leibniz rule
	\be 
	[A,f\,B]=f\,[A,B]+\big(\rho(A)f\big)\,B-\langle A,B\rangle\, {\cal D}f~.
	\ee
	Once more, the last term is absent in the Dorfman formulation and the rule acquires the familiar form of the Leibniz rule for the Lie bracket. However, this is true only for functions in the second slot; due to the non-anti-symmetric nature of the operation, when the function is in the first slot, i.e. $[fA,B]$, the Leibniz rule is still modified. 
Next, the map $\rho$ is a homomorphism,
	\be \rho [A,B]=[\rho(A),\rho(B)]~.
	\ee
The fourth property states that the image of the derivation ${\cal D}$ is in the kernel of the anchor map,
	\be 
	\rho \circ {\cal D}=0~, \quad 
	\mbox{which is equivalent to}
	\quad \langle {\cal D}f, {\cal D}g \rangle=0~.
	\ee
As we will see in the next section, this condition translates into the strong constraint in DFT. 
The final property is the compatibility between the Courant bracket and the symmetric pairing,
	\be 
	\rho(C)\langle A,B\rangle = \langle [C,A]+{\cal D}\langle C,A\rangle,B\rangle + \langle A,[C,B]+{\cal D}\langle C,B\rangle\rangle 
	~.
	\ee
For a minimal set of axioms, one only requires the Jacobi identity, compatibility and any one of the remainder.

To the above five abstract properties, let us introduce a local basis for the sections of $E$, $e^I$ where $I=1,\dots,2d$.
The bilinear form of two basis elements gives the $O(d,d)$ metric,
\be 
\langle e^I,e^J\rangle = \frac{1}{2}\eta^{IJ}
~,
\ee
while their Courant bracket yields
\be [e^I,e^J]=\eta^{IK}\eta^{JL}T_{KLM}e^M 
~,
\ee 
with a generalized three-form $T_{KLM}$.
For operations involving the anchor and the derivation, we use the local coordinate expressions
\be \rho(e^I)f = \eta^{IJ}\rho^i_{\phantom{i}J}\partial_i f
~,
\ee 
\be 
{\cal D}f = {\cal D}_I f \, e^I = \rho^i_{\phantom{i}I}\partial_i f \, e^I
~,
\ee
where the components of the anchor $\rho$ are $(\rho^i_{\phantom{i}J}) = (\rho^i_{\phantom{i}j}, \rho^{ij})$ for $i=1,\dots,d$.
Substituting the local basis in the five abstract properties where we expand $A = A_I e^I$,
the Courant algebroid properties that we can derive in local coordinates
are (see also \cite{Ikeda})
\bea 
\eta^{IJ}\,\rho^i{}_{I}\,\rho^j{}_J &=&0 ~,
\label{Cou_prop_local1}
\\[4pt]
\rho^i{}_{I}\,\partial_i\rho^j{}_J-\rho^i{}_J\,\partial_i\rho^j{}_I-\eta^{KL}\,\rho^j{}_K\,T_{LIJ}&=&0~,\\[4pt]
4\,\rho^i{}_{[L}\,\partial_iT_{IJK]}+3\,\eta^{MN}\,
T_{M[IJ}\,T_{KL]N}&=&0~.
\label{Cou_prop_local3}
\eea
$[....]$ denotes anti-symmetrization with weight 1. Interestingly, the five properties result in only three local coordinate conditions, which is another way to see that only three of them are independent. 
From a physical point of view, the three-form $T_{KLM}$ can be identified with the four types of fluxes $(H_{ijk},f^i_{jk},Q^{ij}_{k},R^{ijk})$ of a general supergravity compactification \cite{Hal09}. In particular, the second equation gives the flux expressions in terms of gauge potentials identified with the components of $\rho$, and the third equation is identical to their Bianchi identities. 

Let us briefly discuss  the correspondence between a
Courant algebroid and a QP2-manifold.
A QP2-manifold is defined by a triplet ($\mathcal{M}=T^*[2]T[1]M, \omega, Q$).
On the manifold, the degree of the cotangent bundle is shifted by two and the degree of the tangent bundle is shifted by one. 
There are a P-structure $\omega$, which is a degree two symplectic structure and a
Q-structure $Q$, which is a degree one vector field.
These structures satisfy the compatibility condition, $\mathcal{L}_Q \omega = 0$.
The Q-structure gives rise to a degree three
{Hamiltonian function} $\Theta$,
$
Q=\{\Theta,\, \cdot \,\}
$
with a Poisson bracket of degree $-2$.
It is cohomological when $Q^2=0$, thus implying the 
{Cartan structure equation}
\be 
\{\Theta,\Theta\}=0 ~, 
\ee
which gives (\ref{Cou_prop_local1}) - (\ref{Cou_prop_local3})
in local coordinates when
\be \label{Th_Cou}
\Theta=\rho^i{}_I(x)\, F_i\,A^I-\sfrac 1{3!}\,T_{IJK}(x)\,A^I\,A^J\,A^K~,
\ee  
where $F_i$, $A^I$ and $x$ are coordinates on
$\mathcal{M}$ of degree two, one and zero, respectively.
Therefore this establishes the correspondence to a canonical Courant algebroid ($E,[\cdot , \cdot]_E, \langle \cdot , \cdot \rangle_E, \rho$) \cite{Roy02}.

\section{An algebroid for double field theory}

Having laid out the properties of a standard Courant algebroid in the previous section,
we now examine the structures of DFT, and whether the above properties are preserved or modified in this case.
In order to make connections from a Courant algebroid to DFT, naturally we would need to double the target space of the canonical Courant algebroid.

As a local model, we consider the target space $T^{\ast}M$
with local coordinates
\be 
\X=(\X^I)=(\X^i,\X_i)=:(X^i,\widetilde X_i)~. 
\ee
The {vector bundle} over this doubled target space $T^{\ast}M$ is 
\be
E=\mathbb T(T^{\ast}M):=T(T^{\ast}M)\oplus T^{\ast}(T^{\ast}M)~.
\ee
Local sections of the bundle  are $(\A^{\hat I})=(\A^I,\widetilde \A_I)=(\A^i,\A_i,\widetilde \A_i,\widetilde \A^i)$, where $\hat I = 1, \dots, 4d$.
Similarly, the generalized vector is composed of a vector and one-form,
$
\A=\A_{V}+\A_{F}:=\A^I\,\partial_I+\widetilde \A_I\,\dd\X^I
$, where
the basis vectors on $T^\ast M$ are
$(\partial_I)=(\partial/\partial X^i,\partial/\partial \widetilde
X_i)=:(\partial_i,\tilde\partial^i)$
and
the basis forms on $T^\ast M$ are
$(\dd\X^I):=(\dd X^i,\dd \widetilde X_i)$. 
Previously in the canonical case, the components of the anchor were $\rho^i{}_{J}$. 
By doubling the target space, indices get doubled accordingly,
so the anchor becomes $\rho^I{}_{\hat{J}} = (\rho^I{}_J,\widetilde\rho^{\,IJ})$.
With this, we define a large Courant algebroid
$(E,[\, \cdot \,,\, \cdot \,],\langle\, \cdot \,,\, \cdot \,\rangle,\rho)$. 
It is simply a canonical Courant algebroid defined over a doubled target space.

Next, we introduce a decomposition on the generalized vector and anchor such that
\be\label{polarise}
\A_{\pm}^I  =\sfrac 12\,\big( \A^I\pm\eta^{IJ}\,\widetilde \A_J\big)~, 
\quad
e^{\pm}_I=\partial_I\pm \eta_{IJ}\,\dd\X^J~,
\ee
and
\bea
(\rho_{\pm})^{I}{}_{J}=\rho^{I}{}_{J}\pm \eta_{JK}\,\widetilde \rho^{\,IK}~.
\eea 
The $O(d,d)$ metric $\eta$ is used, relating their standard and dual parts.
The generalized tangent bundle is thus decomposed as 
\be\label{eq:EL+L-}
E= T(T^{\ast}M)\oplus T^{\ast}(T^{\ast}M) = L_+\oplus L_- \ , 
\ee
where $L_\pm$ is the bundle whose space of sections is spanned locally by $e_I^\pm$. 
The associated {generalized vector} is 
\be 
\label{vec_L+L-}
\A=\A_+^I\,e_I^++\A_-^I\,e_I^-~.
\ee

We would like to recover the DFT structures, namely the DFT vector, C-bracket of the DFT vectors, and the generalized Lie derivative in the theory.
In DFT, one observes that the indices of the fields are 2$d$-dimensional.
The generalized vector $\A^{\hat I}$ with a 4$d$-dimensional index $\hat I$ in the large Courant algebroid is clearly unfit.
Therefore, we need to implement projections.
We introduce a projection to the subbundle $L_+$ of $E$ through the bundle map 
\bea 
\label{proj}
\sfp_+: E \longrightarrow L_+ \ , \quad
(\A_V,\A_F) \longmapsto \A_+:=A~.
\eea 
Under the projection, the components $\A_-^I$
in the generalized vector (\ref{vec_L+L-}) simply vanish, and we rename $\A_+^I=A^I$. 
This leads to a special generalized vector  
\be\label{dftvector}
A=A_i\,\big(\dd X^i+\tilde\partial^i\big)+A^i\,\big(\dd \widetilde X_i+\partial_i\big)~,
\ee
which is a {DFT vector} \cite{DS16}.
Turning to the binary operations, it is useful 
to define an
inclusion map
\be 
\iota : L_+\, \hookrightarrow \, E  
\ee
so that elements of $L_+$ can be treated as elements of $E$.
The inclusion map may be composed with the projection,
\be 
\mathsf{a} := \iota \circ \sfp_+: E \longrightarrow E
~.
\ee
Applying the projection to the standard {Courant bracket} for  elements of $E$ obtained by the action of the map $\mathsf{a}$,
\bea
\sfp_+\big([\mathsf{a}(\A),\mathsf{a}(\B)]_E\big) 
&=& (A^K\,\partial_K B^J-\sfrac 12\, A^K\,\partial^JB_K - \{A\leftrightarrow B\}) e_J^+
 \\
&:=& \cbral A,B\cbrar_{L_+} 
~,
\eea
we obtain the {C-bracket of DFT vectors}.
Similarly, projecting the {Dorfman bracket},
\bea
\sfp_+\big(\L_{\mathsf{a}(\A)} \mathsf{a}(\B)\big) 
&=& (A^I\,\partial_IB^J-B^I\,\partial_IA^J+B_I\,\partial^JA^I)e_J^+
\\
&:=& \sfL_{A}B
~,
\eea
we obtain the {generalized Lie derivative of DFT}. 
Recall that for the gauge transformations to close,
\be 
\label{DFTclosed}
[\sfL_C,\sfL_A] = \sfL_{\cbral
  C,A\cbrar_{L_+}}
  \ee
has to hold, that is the Lie bracket of the DFT generalized Lie derivatives equals the DFT generalized Lie derivative along the respective C-bracket.
This is obtained at the expense of imposing the {strong constraint},
\be 
\eta^{IJ}\,\partial_I f\,\partial_Jg=0
\ee
for all fields $f,g$ of DFT.
In the case of a Courant algebroid, this strong constraint is not needed, as the Dorfman derivative immediately satisfies such a condition as (\ref{DFTclosed}).
Finally, the projection $\sf{p}_{+}$ leads naturally from the $O(2d,2d)$ metric $\hat{\eta}_{\hat{I}\hat{J}}$ and the corresponding bilinear form to the correct $O(d,d)$ structure of DFT, according to
\be 
\langle \sfp_+(\A), \sfp_+(\B) \rangle_{L_+} := \langle \mathsf{a}(\A), \mathsf{a}(\B) \rangle_E
~.
\ee

Now we are ready to outline the definition of an algebroid for DFT.
Let $L_+$ be a vector bundle of
rank $2d$ over $T^{\ast}M$.
There is a skew-symmetric C-bracket, 
$\cbral\, \cdot \,,\, \cdot \,\cbrar_{L_+}:\G(L_+)\otimes\G(L_+)\to\G(L_+)$,
and a non-degenerate symmetric form $\langle
\, \cdot \,,\, \cdot \,\rangle_{L_+}:\G(L_+)\otimes\G(L_+)\to C^\infty(T^*M)$.
Lastly a smooth bundle map $\rho_+:L_+\to T(T^\ast M)$.
To summarize, the quadruple $(L_+,\cbral\, \cdot \,,\, \cdot \,\cbrar_{L_+},\langle
\, \cdot \,,\, \cdot \,\rangle_{L_+},\rho_+)$
defines a {double field theory algebroid}.

The DFT algebroid satisfies three properties. 
The first is related to the strong constraint,
\be 
\label{dft_SC}
\langle {\cal D}_+f,{\cal D}_+g\rangle_{L_+}=\frac{1}{4}\, \langle \dd f,\dd g\rangle_{L_+}
~,
\ee
for functions $f,g\in C^{\infty}(T^*M)$.
The derivative ${\cal D}_+:C^\infty(T^*M)\to\G(L_+)$ is defined through 
\be 
\langle {\cal D}_+f,A\rangle_{L_+}=\sfrac 12\, \rho_+(A)f~.
\ee
Here there is a non-vanishing contribution on the right-hand side of (\ref{dft_SC}),
unlike the case of a Courant algebroid where $\langle {\cal D}f,{\cal D}g\rangle = 0$. 
The second property is the compatibility condition for the C-bracket and the symmetric pairing,		
\bea
\rho_+(C)\langle A,B\rangle_{L_+}
&=&
\langle \cbral C,A\cbrar_{L_+}+{\cal D}_+\langle C,A\rangle_{L_+},B\rangle_{L_+}
%\\ &&
+\langle A,\cbral C,B\cbrar_{L_+}+{\cal D}_+\langle C,B\rangle_{L_+}\rangle_{L_+}
\eea
for $A,B,C\in \G(L_+)$.
The third property is the Leibniz rule,
\be 
\cbral A,f\,B\cbrar_{L_+}=f\,\cbral A,B\cbrar_{L_+}+\big(\rho_+(A)f\big)\,B-\langle A,B\rangle_{L_+}\,{\cal D}_+f
~.
\ee
In comparison with the corresponding Courant algebroid properties, 
we observe that the condition related to the strong constraint is modified in DFT, while the general form of the compatibility and Leibniz properties is not altered. 
This also teaches us that once the first property of a Courant algebroid (modified Jacobi identity) is relaxed, then the previously trivial properties should be examined anew and they indeed do not follow directly. 

More specifically, the homomorphism property gets modified in DFT.
The modification is controlled by the strong constraint,
\be 
\rho_+\cbral A,B\cbrar_{L_+}=[\rho_+(A),\rho_+(B)]+{\sf SC}_\rho(A,B)~,
\ee
where 
\be 
{\sf SC}_\rho(A,B)=\big(\rho_{L[I}\,\partial^K\rho^L{}_{J]}\,A^I\,B^J+\sfrac 12\, (A^I\,\partial^KB_I-B^I\,\partial^KA_I)\big)\,\partial_K~.
\ee 
The ${\sf SC}_\rho$ term vanishes upon imposing the strong constraint,
$\partial^K(\dots) \partial_K (\dots) = 0$.
Similarly the Jacobi identity is satisfied only up to the strong constraint,
\bea 
\cbral\cbral A,B\cbrar_{L_+},C\cbrar_{L_+}+ {cyclic}
&=&{\cal D}_+
{\cal N}_+(A,B,C) + {\cal Z}(A,B,C)
+{\sf SC}_{\rm Jac}(A,B,C)~,
\eea
where 
\be 
{\cal N}_+(A,B,C)=\sfrac 13\, \langle \cbral A,B\cbrar_{L_+},C\rangle_{L_+}+{cyclic}
~,
\ee 
\be 
{\cal Z}_{IJKL} = 4\,\rho^M{}_{[L}\,\partial_{\underline{M}}T_{IJK]}+3\,\eta^{MN}\,
T_{M[IJ}\,T_{KL]N}
~,
\ee 
and
\bea 
{\sf SC}_{\rm Jac}(A,B,C)^L&=&-\sfrac 12\, \big(A^I\,\partial_J B_I\,\partial^JC^L-B^I\,\partial_J A_I\,\partial^JC^L\big)
\nn\\&&
 -\, \rho_{I[J}\,\partial_M
\rho^I{}_{N]}\,\big(A^J\,B^N\,\partial^M C^L
\nn\\&&
-\,\sfrac 12\,
C^J\,A^K\,\partial^M B_K\,\eta^{NL}
%\qquad \qquad \qquad \qquad \qquad 
+\,\sfrac 12\, C^J\,B^K\,\partial^M A_K\,\eta^{NL}\big) 
\nn\\ && 
+\,{cyclic} ~.
\eea
The ${\sf SC}_{\rm Jac}$ term vanishes after imposing the strong constraint.
Here we have also relaxed the Jacobi identity by including a four-form ${\cal Z}_{IJKL}$.
It however also vanishes under the strong constraint.

Let us introduce a local basis for the sections of $L_+$, $e_I$ where $I=1,\dots,2d$,
with the following operations
\be [e_I,e_J]= T_{IJ}^{\phantom{IJ}M}\,e_M~, \quad
\langle e_I,e_J\rangle=\eta_{IJ}~,
\ee
\be 
\rho(e_I)f=\rho^L{}_{I}\,\partial_L f~, \quad
{\cal D}f={\cal D}^I f\,e_I=\sfrac 12 \,\rho^K{}_L\,\partial_Kf\,\eta^{LJ}\,e_{J}
~,
\ee
where the components of the anchor $\rho$ are $(\rho^I_{\phantom{I}J}) = (\rho^i_{\phantom{i}j}, \rho^{ij},\rho_i^{\phantom{i}j}, \rho_{ij})$
for $i=1,\dots,d$.
The pairing carries the signature of the $O(d,d)$ metric $\eta_{IJ}$.
Let us show the first DFT property which is related to the strong constraint in local coordinates,
\be 
\label{SC_rel}
\langle {\cal D}f,{\cal D}g\rangle_{L_+} =
\sfrac 14 (\rho^K{}_{I}\,\eta^{IJ}\,\rho^L{}_J)\,\partial_Kf\,\partial_Lg=\sfrac 14\eta^{KL}\,\partial_Kf\,\partial_Lg~,
\ee
where we have used the fact that
\be 
\label{rho_eta_rho_eta}
\rho^K{}_{I}\,\eta^{IJ}\,\rho^L{}_J = \eta^{KL}~.
\ee
Imposing the vanishing of (\ref{SC_rel}), we regain the Courant algebroid structure.
Overall, the DFT properties in local coordinates are
\bea 
\eta^{IJ}\,\rho^K{}_{I}\,\rho^L{}_J&=&\eta^{KL}~,\\[4pt]
2 \rho^L{}_{[I}\,\partial_{\underline{L}}\rho^K{}_{J]} - \eta^{MN}\,\rho^K{}_M\,\hat{T}_{NIJ}&=& \rho_{L[I}\,\partial^K\rho^L{}_{J]} ~,\\[4pt]
4\,\rho^M{}_{[L}\,\partial_{\underline{M}}\hat{T}_{IJK]}+3\,\eta^{MN}\,
\hat{T}_{M[IJ}\,\hat{T}_{KL]N}&=& \mathcal{Z}_{IJKL}~,
\eea
where the three-form $\hat{T}_{IJK}$ gives the fluxes in DFT \cite{ABMN11, G11, GMNP13, BGHS13}, 
given the components of the anchor as
\be 
\rho^I{}_J= \left(\begin{array}{cc}\d^i{}_j&\beta^{ij}\\ B_{ij}&\d_i{}^j+\beta^{jk}\,B_{ki}\end{array}\right)~.
\ee
$\beta^{ij}$ and $B_{ij}$ are respectively a two-vector and a two-form.
The twist in the C-bracket of DFT $\hat{T}$ is related to the twist in the Courant bracket $T$ by $\hat{T} (A,B) = \sfrac 12 T(A,B)$. Thus we observe that there are once more three conditions, modified with respect to the Courant algebroid ones. Identifying the anchor components with gauge potentials and the twist with the fluxes renders the second expression equivalent to the general fluxes of DFT and the third equation to their Bianchi identities.

To conclude,
in our approach double field theory is a projection from the large Courant algebroid (that is, a Courant algebroid on a doubled target space).
As we have verified based on its properties, the DFT structure indeed reduces to the canonical Courant algebroid under the strong constraint.
The DFT algebroid is later formulated in the graded symplectic geometry in \cite{KSS18}.

\section{Intermediate structures}

In this section we would like to highlight some additional structures. Their appearance is a result of the relaxation of the Courant algebroid axioms.
Here we show the chain of  structures:
\begin{center}
Courant algebroid (CA)
$\stackrel{\cancel{Jacobi}}{\longrightarrow}$ 
Pre-CA
$\stackrel{\cancel{Homomorphism}}{\longrightarrow}$  
Ante-CA
$\stackrel{\cancel{\langle \mathcal{D}f,\mathcal{D}g \rangle=0}}{\longrightarrow}$ 
Pre-DFT algebroid
~.
\end{center}
When one of the five properties of Courant algebroid, namely the Jacobi identity is violated, one gets a structure called pre-Courant algebroid, described originally in \cite{Vais04}.
A pre-Courant algebroid is a quadruple $(E, [\cdot,\cdot],\langle\cdot,\cdot\rangle, \rho)$ which satisfies the properties of a Courant algebroid except for Jacobi.
It is essentially a Courant algebroid twisted by a closed four-form \cite{HS09}, \cite{LSX12}.
In the supermanifold formulation, the structure is called symplectic {\it almost} Lie 2-algeboid \cite{BG16}, where the classical master equation $\{ \Theta,\Theta\} = 0$ is not satisfied and in fact is relaxed to
\be
\{\{\Theta,\Theta\},f\}=0~,
\ee
for any function $f\in C^\infty(M)$, and the degree-three Hamiltonian function
\be \label{Th}
\Theta=\rho^M{}_I(x)\, F_M\,A^I-\sfrac 1{3!}\,T_{IJK}(x)\,A^I\,A^J\,A^K~.
\ee 
In the Hamiltonian, $\rho^M{}_I$ and $T_{IJK}$ are respectively the anchor component and three-form flux that we know of, while
$F_M$ is a degree-two auxiliary field and $A^I$ is a degree-one DFT vector.
The breaking of the Jacobi identity is quantified in
$\{\{\{\{ \Theta, \Theta \}, A\}, B\}, C\}$ for sections of the bundle $A,B,C$.

Similar statements hold for the other structures in the chain.
They are defined by a quadruple $(E, [\cdot,\cdot],\langle\cdot,\cdot\rangle, \rho)$, but satisfying different properties.
An ante-Courant algebroid arises by breaking the Jacobi and homomorphism properties in the Courant algebroid.
In \cite{BG16}, the failure of homomorphism is given by
$\{\{\{\{ \Theta, \Theta \}, f\}, A\}, B\}$ in the supermanifold formalism.
Finally, by relaxing the three properties which are the Jacobi, homomorphism and the strong-constraint-related condition, preserving only the Leibniz and compatibility condition, we obtain a pre-DFT algebroid. 
This structure is known as a symplectic {\it nearly} Lie 2-algebroid
correspondingly in the supermanifold description \cite{BG16}.
On the other hand,
assuming that the Dorfman bracket is equal to the Courant bracket
plus a total derivative,
\be 
\label{Dorf_Cou_brac}
A \circ B = [A,B] + {\cal D} \langle A , B \rangle
~,
\ee
a pre-DFT algebroid is equivalent to a metric algebroid defined originally in \cite{Vaisman12}.
The properties of a metric algebroid are the Leibniz rule
\be 
A \circ (fB) = f (A\circ B) + (\rho(A)f)B
~,
\ee
the compatibility condition
\be 
\rho(C)\langle A , B \rangle = \langle C \circ A,B\rangle + \langle A,C \circ B\rangle
~,
\ee
and the strong-constraint-related condition 
\be 
A\circ A = {\cal D}\langle A , A\rangle
~.
\ee 
The bracket in the metric algebroid $(E,\cdot\circ\cdot,\langle\cdot,\cdot\rangle, \rho)$ is the Dorfman bracket rather than the anti-symmetric Courant bracket.
Therefore this is not an ante-Courant algebroid.
Introducing the Courant bracket as in (\ref{Dorf_Cou_brac}) into the metric algebroid, a pre-DFT algebroid is obtained.

Double field theory is indeed an example of a pre-DFT algebroid, where
the properties are violated simultaneously and in a particular way,
see (\ref{rho_eta_rho_eta}) and thereof, where the homomorphism and Jacobi get modified in a strong-constraint-controlled way.
Therefore,
when imposing the strong constraint, that is,
$\rho^K{}_{I}\,\eta^{IJ}\,\rho^L{}_J  = 0$,
DFT reduces directly to Courant algebroid without encountering any other structures:
\begin{center}
Double field theory 
%$\stackrel{\text{strong constraint}}{\longrightarrow}$ 
$\stackrel{\langle{\cal D}f,{\cal D}g\rangle = 0}{\longrightarrow}$ 
Courant algebroid
~.
\end{center}
This reduction has also been
understood geometrically in \cite{MSS19} as a compatibility condition of
Dirac structures in a metric algebroid.

\acknowledgments
We acknowledge support by COST (European Cooperation in Science and Technology) in the framework of the Action MP1405 QSPACE.
This work was supported by the Croatian Science Foundation under the Project ``New Geometries for Gravity and Spacetime'' (IP-2018-01-7615).
A.Ch., L.J. and F.S.K. were partially supported by
the European Union through the European Regional Development Fund - the Competitiveness and Cohesion Operational Programme (KK.01.1.1.06).
The work of R.J.S. was supported by the Consolidated Grant ST/P000363/1 from the U.K. Science and Technology Facilities Council.

\end{document}